\documentclass[conference]{IEEEtran}
\usepackage{cite}

\ifCLASSINFOpdf
  \usepackage[pdftex]{graphicx}
  \graphicspath{{../figures/}}
  \DeclareGraphicsExtensions{.pdf,.jpeg,.png}
\else
\fi
\usepackage{amsmath}
\usepackage[bookmarks=false,colorlinks=true,linkcolor=blue,urlcolor=cyan]{hyperref}

\usepackage{flushend} 

\begin{document}
%
\title{Integration of physics-derived memristor models with machine learning frameworks}

\author{\IEEEauthorblockN{Zhenming Yu\IEEEauthorrefmark{1}\IEEEauthorrefmark{2}, Stephan Menzel\IEEEauthorrefmark{1}, John Paul Strachan\IEEEauthorrefmark{1}\IEEEauthorrefmark{2}, Emre Neftci\IEEEauthorrefmark{1}\IEEEauthorrefmark{2}}

\IEEEauthorblockA{\IEEEauthorrefmark{1}Fakultät für Elektrotechnik und Informationstechnik, RWTH Aachen, Aachen, 52074, Germany}
\IEEEauthorblockA{\IEEEauthorrefmark{2}Peter Grünberg Institut, Forschungszentrum Jülich GmbH, Jülich, 52425, Germany}
\IEEEauthorblockA{\{z.yu, e.neftci\}@fz-juelich.de}}


%


\maketitle

\begin{abstract}



Simulation frameworks such MemTorch \cite{lammie2020memtorch}\cite{lammie2022memtorch}, DNN+NeuroSim \cite{peng2020dnn+}\cite{yu2021compute}, and aihwkit \cite{rasch2021flexible} are commonly used to facilitate the end-to-end co-design of memristive machine learning (ML) accelerators. These simulators can take device nonidealities into account and are integrated with modern ML frameworks. However, memristors in these simulators are modeled with either lookup tables or simple analytic models with basic nonlinearities. These simple models are unable to capture certain performance-critical aspects of device nonidealities. For example, they ignore the physical cause of switching, which induces errors in switching timings and thus incorrect estimations of conductance states. 
This work aims at bringing physical dynamics into consideration to model nonidealities while being compatible with GPU accelerators. We focus on Valence Change Memory (VCM) cells, where the switching nonlinearity and SET/RESET asymmetry relate tightly with the thermal resistance, ion mobility, Schottky barrier height, parasitic resistance, and other effects \cite{cuppers2019exploiting}. The resulting dynamics require solving an ODE that captures changes in oxygen vacancies. We modified a physics-derived SPICE-level VCM model \cite{bengel2020variability}\cite{ntinas2022towards}, integrated it with the aihwkit \cite{rasch2021flexible} simulator and tested the performance with the MNIST dataset. Results show that noise that disrupts the SET/RESET matching affects network performance the most. This work serves as a tool for evaluating how physical dynamics in memristive devices affect neural network accuracy and can be used to guide the development of future integrated devices.

\end{abstract}


%
\IEEEpeerreviewmaketitle

\section{Introduction}
\label{sec:intro}
Because of their compact size, non-volatility, and low latency, memristive devices show great potential in ML and neuromorphic engineering. Digital, analog, and stochastic in-memory computing schemes have been developed that utilize the advantages of memristors~\cite{ielmini2018memory}. However, memristors are subject to nonidealities like switching nonlinearity, SET/RESET asymmetry, device-to-device, and cycle-to-cycle variations. Naive training algorithms that don't take these into account suffer from performance loss.~\cite{lee2022impact} To assist in co-designing memristive ML accelerators, simulation frameworks have been developed~\cite{lammie2021modeling} with various memristor models.

Memristor models can be generally sorted into two categories: behavioral models and physics-derived models. With behavioral models, memristors are treated as black boxes. Experimental observations are fitted with simple equations, and the models are validated and improved in this process. In contrast, physics-derived models formulate physical equations stemming from an analysis of physical phenomena. The derived equations are often simplified and optimized to get the final solution. While physics-derived models that can accurately produce voltage-dependent behaviors have been adopted in circuit design and validations~\cite{fu2022high}\cite{mayahinia2022voltage}, only behavioral models were used in ML simulators like MemTorch \cite{lammie2020memtorch}\cite{lammie2022memtorch}, DNN+NeuroSim \cite{peng2020dnn+}\cite{yu2021compute}, and aihwkit \cite{rasch2021flexible}.

Behavioral models can produce faithful results at a rather low compute cost, but they are not able to capture some aspects of device physics, which limits the application of ML simulators. For example, they do not model voltage-dependent switching behaviors, so the effect of different SET and RESET voltages cannot be investigated and optimized. They do not model noise based on variations of device parameters, so the simulation results cannot be used to guild material scientists for optimizing memristive devices. To improve on this situation, we integrated a physics-derived memristor model in an open-source ML simulator aihwkit \cite{rasch2021flexible} and investigated the effect of device nonidealities. With this approach, we provide a easy access to the voltage configurations as well as physic driven noises, and make it easy to investigate their impact on the network.
 

\section{Device Modeling}
\label{sec:model}
\subsection{JART VCM Model}
\label{sec:jart}
\begin{figure}[!tb]
\centering
\includegraphics[width=2.5in]{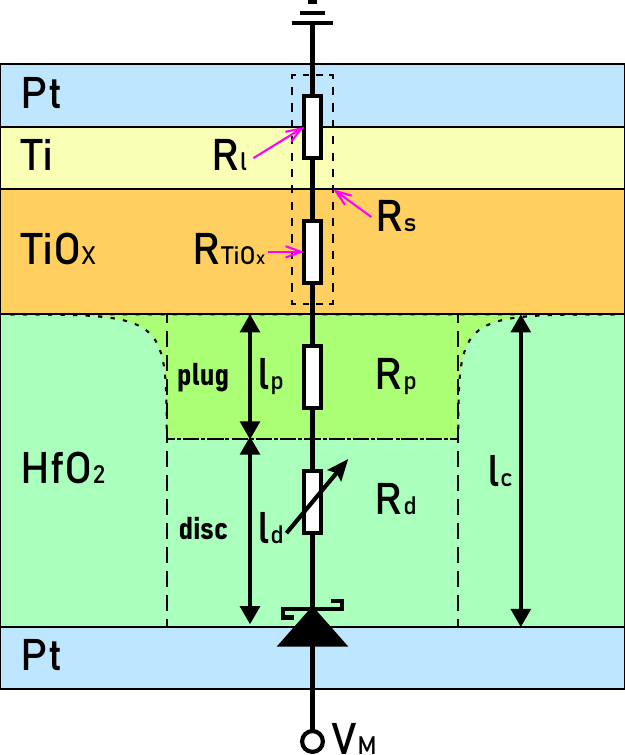}
\caption{Equivalent circuit diagram for the JART memristor model. Details of this device can be found in\cite{hardtdegen2018improved}}
\label{fig:jart_diagram}
\end{figure}

We choose an accurate physics driven model, the Jülich Aachen Resistive Switching Tools (JART) VCM model~\cite{cuppers2019exploiting} as our starting point. The model includes device-to-device and cycle-to-cycle noise, and has also been validated with experimental results in \cite{bengel2020variability}. In the JART model, the filament region of VCM devices are abstracted as a stack of different materials, and the layers of the stack as then modeled as circuit elements in series. As shown in Fig.~\ref{fig:jart_diagram}, the model equivalent circuit consists 4 different parts: $R_s$, $R_p$, $R_d$ and $D_{Sch}$.

$R_s$ represents the series resistance, which consists of a fixed resistance of the titanium oxide layer $R_{TiO_x}$ and a current-dependent line resistance $R_l$. The detailed expression is shown as:
\begin{equation}
\label{eq:Rs}
    R_s=R_{TiO_x}+R_l=R_{TiO_x}+R_0\left(1+\alpha_lR_0{I_M}^2R_{th,l}\right)
\end{equation}
where $R_0$ is the line resistance under zero current, $\alpha_l$ is the temperature coefficient of the lines, $I_M$ is the current through the memristor and $R_{th,l}$ is the thermal resistance of the lines.

The ${\text{HfO}}_{\text{2}}$ layer is divided into two regions, a plug region with fixed oxygen vacancy concentration $N_p$, and a disc region where the oxygen vacancy concentration $N_d$ can change. The conductive plug region serves as a reservoir of oxygen vacancies, which can then flow into or out of the disc region under the applied voltages, and change the conductance. The resistances of these layers are expressed as:
\begin{equation}
\label{eq:Rpd}
    R_{p/d}=l_{p/d}\left(Z_{V_O}eAN_{p/d}\mu_n\right)
\end{equation}
where $l_{p/d}$ is the length of the regions, $Z_{V_O}e=2e$ is the charge of a oxygen vacancy, $A=\pi {r_d}^2$ is the cross section area of the filament area and $\mu_n$ is the electron mobility.

Finally, $D_{Sch}$ represents the Schottky barrier formed between the disc region and the bottom electrode. The current-voltage relation of the resulting Schottky diode is shown as:
\begin{equation}
\label{eq:DSch}
  I_{M} =
    \begin{cases}
      -\sqrt{\pi W_{00}e\left(\frac{\phi_{Bn}}{\cosh^2{\left(\frac{W_{00}}{k_BT}\right)}}-V_{Sch}\right)}\cdot\exp\left(\frac{-e\phi_{Bn}}{W_0}\right)\\
      \cdot\left(\exp\left(\frac{-eV_{Sch}}{{\epsilon'}}\right)-1\right)\cdot AA^*\cdot\frac{T}{k_B}
      \hfill \text{if $V_{M}<0$}\\
      \\
      AA^*T^2\exp\left(\frac{-e\phi_{Bn}}{k_BT}\right)\exp\left(\frac{eV_{Sch}}{k_BT}-1\right)
      \hfill \text{if $V_{M}>0$}
    \end{cases}       
\end{equation}
where $k_B$ is the Boltzman's constant, $A^*$ is the Richardson's constant. The local temperature $T$ is given by:
\begin{equation}
\label{eq:T}
  T=
    \begin{cases}
      I_M\left(V_d+V_p+V_{Sch}\right)R_{th,SET}+T_0
      \hfill \text{if $V_{M}<0$}\\
      I_M\left(V_d+V_p+V_{Sch}\right)R_{th,RESET}+T_0\ \ \ 
      \hfill \text{if $V_{M}>0$}
    \end{cases}     
\end{equation}
where $T_0$ is the ambient temperature, $R_{th,SET}$ and $R_{th,RESET}$ are the thermal resistances of the hafnium oxide layer, and $W_{00}$ and $W_{0}$ are expressed as:
\begin{equation}
\label{eq:W00}
    W_{00}=\frac{eh}{4\pi}\sqrt{\frac{Z_{V_O}N_d}{m^*\epsilon}},\ \ \  W_0=\frac{W_{00}}{\tanh{\left(\frac{W_{00}}{k_BT}\right)}}
\end{equation}
and $\epsilon'$ as:
\begin{equation}
\label{eq:epsilon'}
    \epsilon'=\frac{W_{00}}{\frac{W_{00}}{k_BT}-\tanh{\left(\frac{W_{00}}{k_BT}\right)}}
\end{equation}
where $h$ is the Planck's constant, $m^*$ is the electron effective mass and $\epsilon$ is the static oxide permittivity. The Schottky barrier height $\phi_{Bn}$ is given by:
\begin{equation}
\label{eq:phiBn}
    \phi_{Bn}=\phi_{Bn0}-\sqrt[\leftroot{-4}\uproot{16}4]{\frac{Z_{V_O}N_d\left(\phi_{Bn0}-\phi_n-\frac{V_{Sch}}{e}\right)}{8\pi^2\epsilon^3_{\phi_B}}}
\end{equation}
where $\phi_{Bn0}$ is the nominal Schottky barrier height, $\phi_n$ is the energy level difference between the Fermi level and the conduction band edge and $\epsilon_{\phi_B}$ is the hafnium oxide permittivity related to energy barrier lowering.

\subsection{Simplified Model}
\label{sec:simplified}
The JART model described in Sec.~\ref{sec:jart} is accurate but expensive to compute. The current $I_M$ described in Eq.~\ref{eq:DSch} is only related to the voltage drop across the Schottky diode $V_{Sch}$. However, other parts of the device are modeled as resistors, where the voltage drop can only be calculated with the current $I_M$. Together with other effects, Eq.~\ref{eq:Rs}-\ref{eq:phiBn} form a complex set of nonlinear equations that can only be solved iteratively, which is very hard to parallelize. To solve this issue, we simplified the current calculation with a model~\cite{ntinas2022towards} that is mathematically fitted to the JART model. The simplified model provides an estimation of $I_M$ with the oxygen vacancy concentration in the disc region $N_d$ and the applied voltage across the memristor $V_M$. If $V_{M}<0$:
\begin{equation}
\label{eq:simplified_neg}
\begin{split}
    I_M=-a-\frac{b}{\left(1+c^d\right)^f}\ \ wher&e\ \ 
    a=\frac{a_1+a_0}{1+e^{-\frac{V_M+a_2}{a_3}}}-a_0\\
    b=b_1\left(1-e^{-V_M}\right)-b_0V_M\ \ &,\ \ 
    c=\frac{c_2e^{-\frac{V_M}{c_3}}+c_1V_M-c_0}{N_d\cdot10^{-26}}\\
    d=d_2e^{-\frac{V_M}{d_3}}+d_1V_M-d_0\ \ &,\ \ 
    f=f_0+\frac{f_1-f_0}{1+\left(\frac{-V_M}{f_2}\right)^{f_3}}
\end{split}
\end{equation}
where $a_x$, $b_x$, $c_x$, $d_x$ and $f_x$ are fitting coefficients.
If $V_{M}>0$:
\begin{equation}
\label{eq:simplified_pos}
    I_M=-\frac{g_0\left(e^{-g_1V_M}-1\right)}{\left(1+\left(h_0+h_1V_M+h_2e^{-h_3V_M}\right)\left(\frac{N_d}{N_d,min}\right)^{-j_0}\right)^{\frac{1}{k_0}}}
\end{equation}
where $g_x$, $h_x$, $j_x$ and $k_x$ are fitting coefficients.

Using this simplified model, we can directly calculate $I_M$, and assign the voltage drop across the resistance layers with Eq.~\ref{eq:Rs}-\ref{eq:Rpd}. The $V_{Sch}$ can then be calculated as:
\begin{equation}
\label{eq:VSch}
    V_{Sch}=V_M-V_s-V_p-V_d
\end{equation}
The updates can then be calculated with these voltages.
\subsection{Conductance Update}
\label{sec:update}
The conductance update is calculated with the original JART model~\cite{cuppers2019exploiting}\cite{bengel2020variability}. The conductance change is caused by changing oxygen vacancy concentration in the disc region $N_d$, which is described by an ordinary differential equation:
\begin{equation}
\label{eq:dNdt}
    \frac{dN_d}{dt}=-\frac{I_{ion}}{Z_{V_O}eAl_d}
\end{equation}
where the ionic current $I_{ion}$ is given by:
\begin{multline}
\label{eq:Iion}
    I_{ion}=Z_{V_O}eAc_{V_O}\alpha\nu_0F_{limit}\left(N_d\right)\\
    \cdot\left(\exp{\left(-\frac{\Delta W_{A_{f}}}{k_BT}\right)}-\exp{\left(-\frac{\Delta W_{A_{r}}}{k_BT}\right)}\right)
\end{multline}
$\alpha$ is the ion hopping distance, $\nu_0$ is the attempt frequency and $c_{V_O}$ is the average vacancy concentration:
\begin{equation}
\label{eq:cvo}
    c_{V_O}=\frac{N_p+N_d}{2}
\end{equation}
Function $F_{limit}\left(N_d\right)$ scales the update, and limits $N_d$ within the specified range between $N_{d,max}$ and $N_{d,min}$:
\begin{equation}
\label{eq:Flimit}
  F_{limit}\left(N_d\right) =
    \begin{cases}
      1-\left(\frac{N_d}{N_{d,max}}\right)^{10}\ \ \ \ \ 
      \hfill \text{if $V_{M}<0$}\\
      1-\left(\frac{N_{d,min}}{N_d}\right)^{10}\ \ \ \ \ 
      \hfill \text{if $V_{M}>0$}
    \end{cases}.      
\end{equation}
And the energy barriers for ion hopping are described as:
\begin{equation}
\label{eq:dWA}
    \Delta W_{A_{f/r}}=\Delta W_A\left(\sqrt{1-\gamma^2}\mp \gamma\frac{\pi}{2}+\gamma\arcsin{(\gamma)}\right)
\end{equation}
\begin{equation}
\label{eq:gamma}
    \gamma=\frac{eZ_{V_O}\alpha E_{ion}}{\Delta W_A\pi}
\end{equation}
\begin{equation}
\label{eq:Eion}
    E_{ion}=
    \begin{cases}
      V_d/l_d\ \ \ \ \ 
      \hfill \text{if $V_{M}<0$}\\
      \left(V_p+V_d+V_{Sch}\right)/l_c\ \ \ \ \ 
      \hfill \text{if $V_{M}>0$}
    \end{cases}     
\end{equation}
where $\Delta W_A$ is the activation energy, $l_d$ is the thickness of the disc region and $l_c$ is the thickness of the hafnium oxide layer. 

\begin{figure}[!htb]
\centering
\includegraphics[width=\hsize]{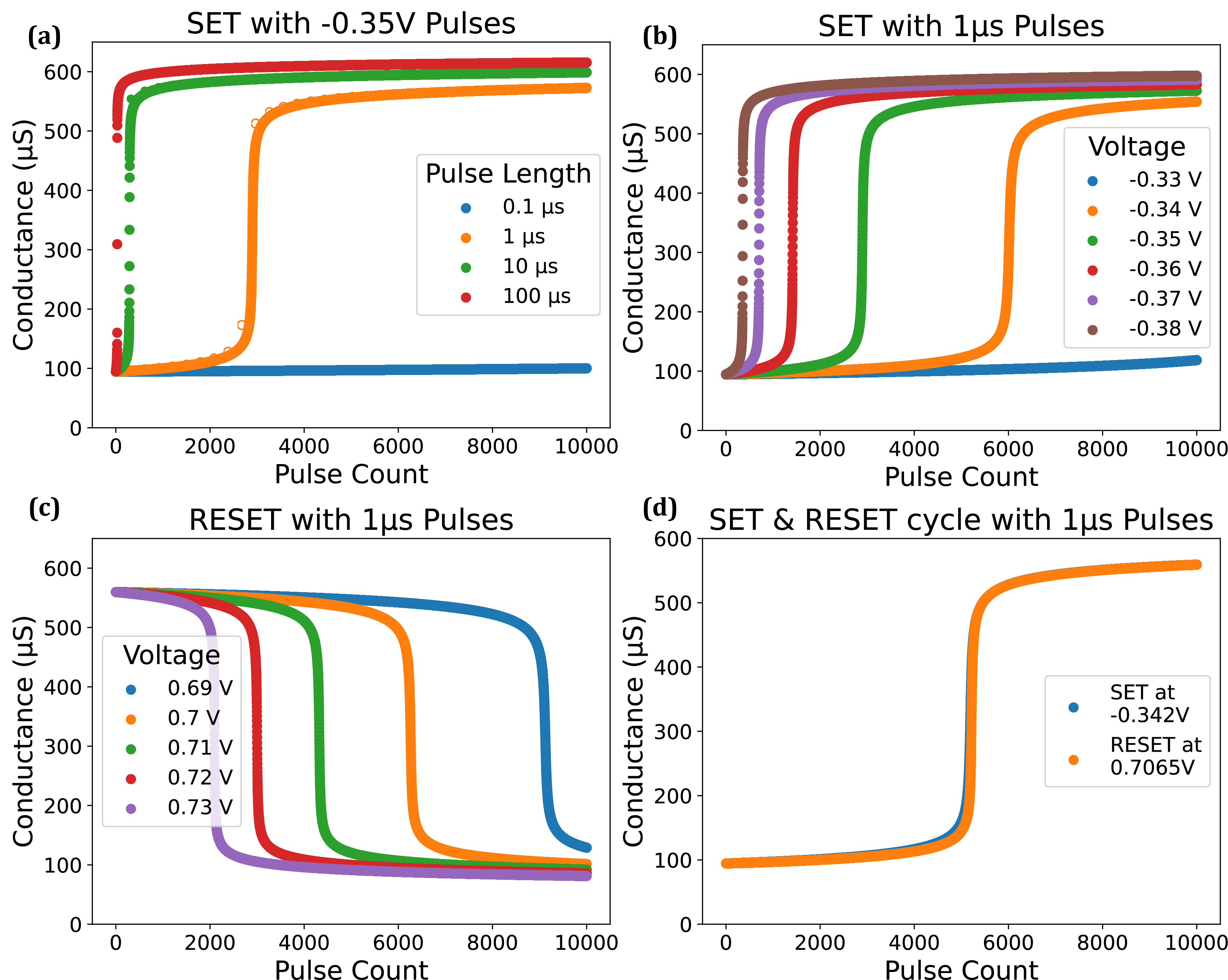}
\caption{\textbf{Simulation results of the memristor model in Python.} \textbf{(a)}: Changing pulse length with fixed pulse amplitude. \textbf{(b)}: Changing pulse amplitude with fixed pulse length for the SET direction(i.e. $V_M<0$). \textbf{(c)}: Changing pulse amplitude with fixed pulse length for the RESET direction(i.e. $V_M>0$). \textbf{(d)}: Hand-tuned SET and RESET matching.}
\label{fig:switching_curves}
\end{figure}

With this physics-derived model, we can freely adjust the pulse length and the pulse amplitude to get various switching behaviour in both directions. As shown in Fig.~\ref{fig:switching_curves}(a-c), higher pulse amplitudes and longer pulse length can make the device switch faster, but the intermediate states becomes less accessible. Lower pulse amplitudes can give us better accessibility to the intermediate states, but at the cost of switching delays. We matched the SET and RESET curves by hand tuning the voltages, and the results shown in Fig.~\ref{fig:switching_curves}(d) are almost identical.

To ease the use of the device in a on-chip learning scenario where the weights are first trained for task A and later adjusted for task B, we need to control the conductance range. If the conductance range is not controlled, a device might be pushed too far off towards the extreme. In this situation, it which will be hard to move back into the fast-switching region, with the low voltages that grant us access to the intermediate states. This could be implemented by control circuits that checks the device conductance, and skip some of the applied pulses that further push deice conductance beyond the specified range.

\subsection{Noise Implementation}
\label{sec:noise}
To account for stochasticity, we implement variations selectively on parameters that are physically noisy. However, as described in Sec.~\ref{sec:simplified}, in the simplified model, physical values are hidden behind the fitting parameters $a_x$-$k_x$ in Eq.~\ref{eq:simplified_neg}-\ref{eq:simplified_pos}. So we only induced noise with conductance update described in Sec~\ref{sec:update}. In the original JART model~\cite{bengel2020variability}, noise are introduced in the oxygen vacancy boundaries: $N_{d,max}$, $N_{d,min}$ and the filament geometry parameters: $r_d$, $l_d$. Because we introduced peripheral circuits that control the conductance ranges, we also added device-to-device noise on the conductance boundaries set by the control circuits: $G_{max}$, $G_{min}$.

In \cite{bengel2020variability}, device-to-device noise are introduced upon initialization, by drawing from a Gaussian distribution with variance scaled by the mean:
\begin{equation}
\label{eq:d2dnoise}
    X_{d2d}\sim \mathcal{N}\left(X_{mean}, X_{mean}\cdot\sigma\right)
\end{equation}
and cycle-to-cycle noise is implemented as a random walk starting from the device-to-device initialization values, i.e. $X_{0} = X_{d2d}$. For $N_{d,max}$ and $N_{d,min}$, cycle-to-cycle noise were directly inserted:
\begin{equation}
\label{eq:c2cNd}
    X_{t+1}=X_t+\Omega\cdot X_t\cdot\sigma,\ \ \ \Omega\sim U\left(-1,1\right)
\end{equation}
and for the filament geometry parameters $r_d$, $l_d$, the cycle-to-cycle noise is scaled by the update magnitude. This produces more noise when a larger update is applied, and less noise otherwise. However, with our ML use case, most weight updates occur in the first few epochs. As learning progresses, the updates become small and the cycle-to-cycle noise becomes negligible. So we added another additive cycle-to-cycle noise to the JART implementation:
\begin{equation}
\begin{split}
\label{eq:c2cGeometry}
X_{t+1}=
    \begin{cases}
      X_t\cdot\left(1+\Omega_1\sigma_{add}+\Omega_2\sigma_{mult}\left(\frac{N_d-N_{d,old}}{N_{d,max}-N_{d,old}}\right)\right)\\
      \hfill \text{if $V_{M}<0$}\\
      X_t\cdot\left(1+\Omega_1\sigma_{add}+\Omega_2\sigma_{mult}\left(\frac{N_{d,old}-N_d}{N_{d,old}-N_{d,min}}\right)\right)\\
      \hfill \text{if $V_{M}>0$}
    \end{cases}\\
    \Omega_1,\ \Omega_2\sim U\left(-1,1\right).
\end{split}
\end{equation}
This additive noise accounts for the diffusion in the filament area, which happens with or without external stimulation.

Optional boundaries can be specified to truncate the device-to-device noise to reasonable ranges or to limit the cycle-to-cycle random walk process during training.


\section{Results}
\label{sec:results}
We integrated the model described in Sec.~\ref{sec:model} with the IBM aihwkit simulator~\cite{rasch2021flexible} and tested the results on MNIST dataset~\cite{LeCun2005TheMD}\footnote{The code for the modified aihwkit simulator with JART model integration and the test scripts used for this work are publicly available on GitHub at \href{https://github.com/ZhenmingYu/aihwkit}{https://github.com/ZhenmingYu/aihwkit}.}. The model is integrated into a 3-layer fully connected network with sigmoid activation between the layers. We trained the network with stochastic gradient descent and a scheduler that decreases the learning rate by $50\%$ for every 10 epochs. This is necessary because our device model has a very sharp transition in the middle of the conductance range (Fig.\ref{fig:switching_curves}). If a constant learning rate is used, the network parameters oscillate towards the end of the training, which damages the performance.

\begin{figure}[!htb]
\centering
\includegraphics[width=\hsize]{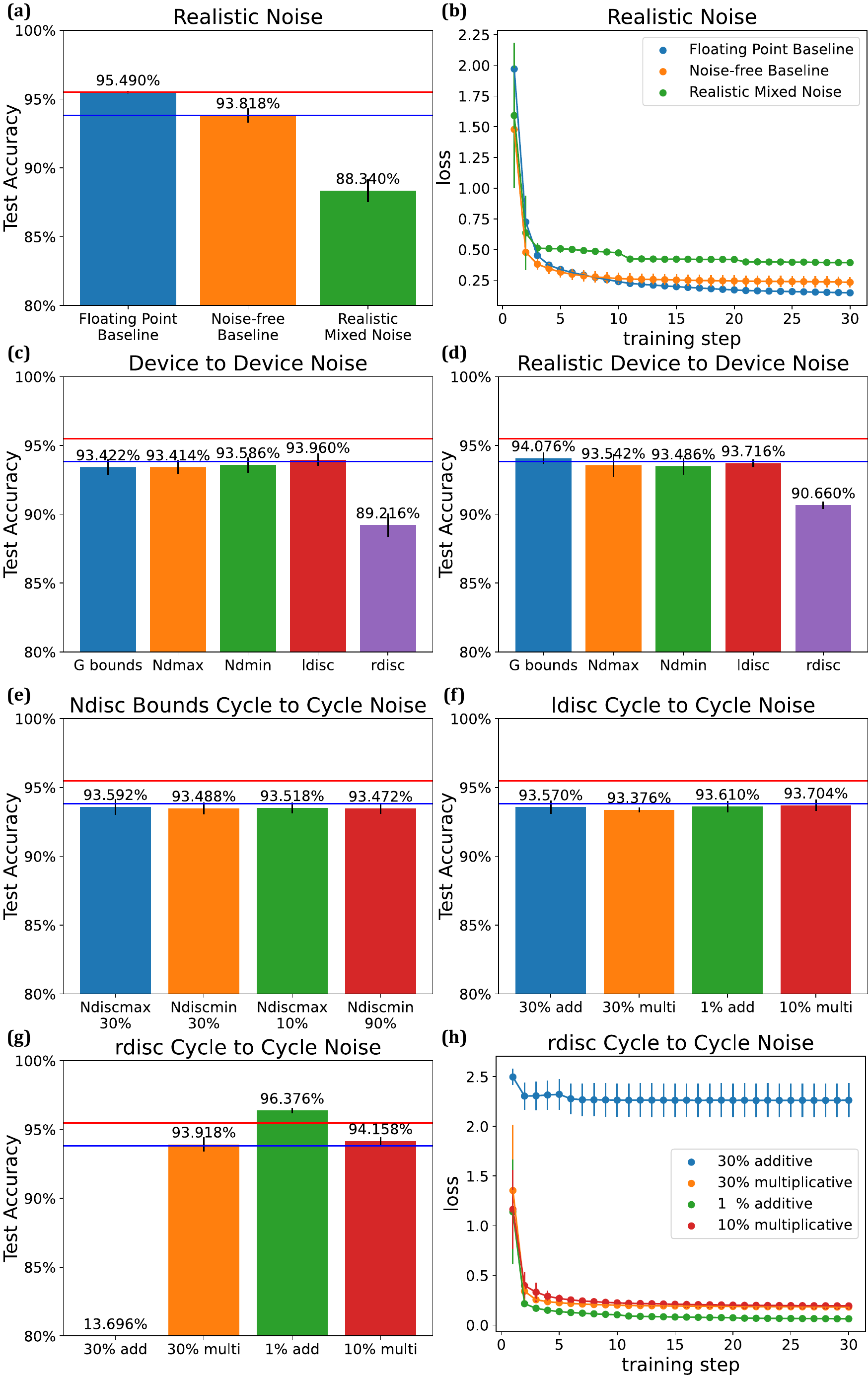}
\caption{\textbf{MNIST simulation results of} accuracy \textbf{(a)} and loss \textbf{(b)} for simulations with realistic mixed noise, without any noise, and with floating point weights. Accuracy of device-to-device noise at the same scale \textbf{(c)} and based on realistic estimation \textbf{(d)}. Accuracy of cycle-to-cycle noise on $N_{d,max}$, $N_{d,min}$ \textbf{(e)} and on $l_{d}$. Accuracy \textbf{(f)} and loss \textbf{(g)} for cycle-to-cycle noise on $r_{d}$.}
\label{fig:results}
\end{figure}

As shown in Fig.~\ref{fig:results}(a), with the same network architecture, learning rate, and scheduler configuration, floating point weights achieved an accuracy of $95\%$. Our device model achieved an accuracy of $93.8\%$ without any noise being applied, and with noise fitted experimentally in~\cite{bengel2020variability}, we can get a performance of $88.3\%$.

From equations~\ref{eq:dNdt}-\ref{eq:Eion}, we can derive that noise on different parameters have different effects. Noise on $N_{d,max}$, $N_{d,min}$, and $l_{d}$ only affect the update magnitude, they don't change the sign of the update, so in the end they only affect the learning rate. Noise on $r_{d}$ however, can affect the matching between the increment and decrement update steps. This is mainly due to the asymmetric change in $E_{ion}$ and $T$ with variable $r_{d}$. As shown in Eq.~\ref{eq:Eion}, with SET pulses, $E_{ion}$ is related with $r_{d}$:
\begin{equation}
\label{eq:Eion_SET}
\begin{split}
    E_{ion}&=\frac{V_d}{l_d}=\frac{I_M\cdot R_d}{l_d}\\
    &=\frac{I_M\cdot l_d\left(Z_{V_O}eAN_d\mu_n\right)}{l_d}\\
    &=I_MZ_{V_O}eN_d\mu_n\pi r_d^2.
\end{split}
\end{equation}
However, with RESET pulses, $E_{ion}$ is only related to noise-free parameters:
\begin{equation}
\label{eq:Eion_RESET}
\begin{split}
    E_{ion}&=\frac{V_p+V_d+V_{Sch}}{l_c}=\frac{V_M-V_s}{l_c}\\
    &=\frac{V_M-I_M\cdot \left(R_{TiO_x}+R_0\left(1+\alpha_lR_0{I_M}^2R_{th,l}\right)\right)}{l_c}.
\end{split}
\end{equation}
Noise in $r_{d}$ also affects the local temperature $T$ through thermal resistances, and Eq.~\ref{eq:T} becomes:
\begin{equation}
\label{eq:T_noisy}
  T=
    \begin{cases}
      I_M\left(V_M-V_s\right)R_{th,SET}\cdot\frac{r_d^2}{r_{d,Noisy}^2}+T_0
      \hfill \text{if $V_{M}<0$}\\
      I_M\left(V_M-V_s\right)R_{th,RESET}\cdot\frac{r_d^2}{r_{d,Noisy}^2}+T_0\ 
      \hfill \text{if $V_{M}>0$}
    \end{cases}.     
\end{equation}
where $R_{th,SET}$ and $R_{th,RESET}$ are different values. As a result, the local temperature will scale differently during SET and RESET for the same $r_{d,Noisy}$, and will then affect the switching behaviour in different ways.

The effect of different noise can also be seen in the results. Fig.~\ref{fig:results}(c) and (d) show the accuracy of the network with device-to-device noise. When the device-to-device noise are scaled to the same extent of $30\%$, noise on most parameters don't affect the performance as much, and the network achieved accuracy close to the noise-free baseline. With device-to-device noise on $r_d$ however, the network performance decreases significantly to about $89.2\%$. Using $r_d$ noise based on a realistic estimation, where $r_d$ is allowed a smaller spread, the performance is slightly increased to about $90.7\%$. Similar results can be seen with cycle-to-cycle noise. As shown in Fig.~\ref{fig:results}(e) and (f), noise on $N_{d,max}$, $N_{d,min}$ and $l_{d}$ don't affect performance much. The results are quite different for the cycle-to-cycle noise on $r_d$: The network is crippled with a large $30\%$ noise directly applied in the random walk. The loss will not go down during training (Fig.~\ref{fig:results}(h)), and an accuracy of only $13.7\%$ is achieved (Fig.~\ref{fig:results}(g)). However, with a small $1\%$ noise directly applied, it becomes easier for the network to get out of local minima, and an accuracy beyond the baseline is achieved. Multiplicative noise scaled by the update have very little impact on the network performance, as they become negligible after a few training epochs.

Because we used the simplified fit model (Sec.~\ref{sec:simplified}) for current calculation, the voltage estimation across different layers will not change with noise on the disc region length $l_{d}$. As a result, noise on $l_{d}$ appear to have little impact on the network performance in our tests. However, in reality, $l_{d}$ should also be able to affect the network by changing the voltage drop across different layers, as shown in Eq.~\ref{eq:Rs}-\ref{eq:DSch}. Nonetheless, we are still able to conclude that the matching between increment and decrement update steps is the most important property for training neural networks on memristors. This finding is also in line with previous publication~\cite{lee2022impact}\cite{gokmen2020algorithm}.

\section{Conclusion}
We built a memristive machine learning simulator with physics-derived model and investigated the effect of noise from various sources based on it. 
This simulator unlocks many possibilities for the community. It opens up access to the pulse configurations, which enables us to explore better ways to match the performance-critical SET and RESET behaviours. It also exposes nonidealities rooted in different physical sources, which grants the ML community chances to develop new algorithms that target these nonidealities directly. On the other hand, it also gives material scientists a tool to evaluate the effect of different noise sources, which can guild them in developing novel device structures that alleviates noise on performance-critical parameters. With the random walk based cycle-to-cycle noise implementation, we can also disable the boundaries and train the model on different tasks. The parameters would then stray away from their original values, causing the device behavior to change after the training. This allow us to study the effect of device aging, and try to prevent performance loss while operating the devices.


 \section*{Acknowledgment}
This work was sponsored by the Federal Ministry of Education, Germany (project NEUROTEC-II grant no. 16ME0398K and 16ME0399). We thank Vasileios Ntinas for his help with the simplified model~\cite{ntinas2022towards}, and Malte J. Rasch for his generous support on aihwkit~\cite{rasch2021flexible}.

 \section*{Copyright}
© 2022 IEEE. Personal use of this material is permitted. Permission from IEEE must be obtained for all other uses, in any current or future media, including reprinting/republishing this material for advertising or promotional purposes, creating new collective works, for resale or redistribution to servers or lists, or reuse of any copyrighted component of this work in other works.


\bibliographystyle{IEEEtran}
\bibliography{IEEEtran/citations}

\end{document}